\documentclass[reprint,prd,twocolumn,superscriptaddress,nofootinbib]{revtex4-1}

\usepackage{color}
\usepackage{amsmath}
\usepackage{graphicx}
\usepackage{scalefnt}

\usepackage{grffile} 

\graphicspath{ {figs_ve/} }

\allowdisplaybreaks

\begin{document}
\hfill TTP22-009,  P3H-22-016
\title{Massive vector form factors to three loops}
\author{Matteo Fael}
\email{matteo.fael@kit.edu}
\affiliation{Institut f\"ur Theoretische Teilchenphysik,
  Karlsruhe Institute of Technology (KIT), 76128 Karlsruhe, Germany}
\author{Fabian Lange}
\email{fabian.lange@kit.edu}
\affiliation{Institut f\"ur Theoretische Teilchenphysik,
  Karlsruhe Institute of Technology (KIT), 76128 Karlsruhe, Germany}
\affiliation{Institut f{\"u}r Astroteilchenphysik,
  Karlsruhe Institute of Technology (KIT), 76344 Eggenstein-Leopoldshafen, Germany}
\author{Kay Sch\"onwald}
\email{kay.schoenwald@kit.edu}
\affiliation{Institut f\"ur Theoretische Teilchenphysik,
  Karlsruhe Institute of Technology (KIT), 76128 Karlsruhe, Germany}
\author{Matthias Steinhauser}
\email{matthias.steinhauser@kit.edu}
\affiliation{Institut f\"ur Theoretische Teilchenphysik,
  Karlsruhe Institute of Technology (KIT), 76128 Karlsruhe, Germany}

\begin{abstract}
  We compute the three-loop non-singlet corrections to the photon-quark form
  factors taking into account the full dependence on the virtuality of the
  photon and the quark mass.  We combine the method of differential
  equations in an effective way with expansions around regular and singular
  points. This allows us to obtain results for the form factors with an
  accuracy of about eight to twelve digits in the whole kinematic range.
\end{abstract}

\maketitle


\bigskip {\bf Introduction.}  Form factors are fundamental objects in Quantum
Chromodynamics (QCD) with a variety of applications.  On the one hand, they
are the simplest objects which show a non-trivial infrared struture and thus
form factors are often used to develop and test all-order theorems about the
infrared singularities of scattering amplitudes in QCD (see, e.g.,
  Refs.~\cite{Mitov:2006xs,Becher:2009kw,Beneke:2009rj}).  On the other hand,
form factors play a crucial role as building blocks in a number of observables
which range from low energies to cross sections at the Large Hadron Collider
(LHC) at CERN.  They describe the universal structure of the
$(Z^\star, \gamma^\star) \to \bar{Q} Q$ vertex function, involving two
on-shell quarks $Q$ and vector or axial-vector couplings of the vector bosons.
Massive form factors enter several processes involving heavy quarks at hadron
and $e^+e^-$ colliders, such as $\bar{t}t$
production~\cite{Moch:2008qy,Kidonakis:2008mu,Moortgat-Pick:2015lbx} and gauge
and Higgs boson
decays~\cite{Bernreuther:2005gw,Bernreuther:2018ynm,Behring:2019oci}, which
clearly require the inclusion of mass effects.  Such processes can probe
deviations of the quark couplings from their values in the Standard
Model. Form factors contribute to the all-virtual corrections to cross
sections.

In Quantum Electrodynamics (QED) lepton masses are often kept to regulate
collinear singularities. Therefore massive form factors take part also in the
differential cross section of low-energy lepton scatterings as for instance
the elastic $e$-$p$ scattering~\cite{Bucoveanu:2018soy,Banerjee:2020rww}, one
of the main avenues for proton radius
measurements~\cite{A1:2010nsl,Xiong:2019umf}, or the $\mu$-$e$
scattering~\cite{CarloniCalame:2020yoz,Banerjee:2020rww}, a process able to
determine the leading hadronic contribution to the muon anomalous magnetic
moment~\cite{CarloniCalame:2015obs,Abbiendi:2016xup,loimuone,Banerjee:2020tdt}.

For massless quarks three-loop corrections to the photon-quark form factor
have been computed more than 10 years ago~\cite{Baikov:2009bg} (see also
Refs.~\cite{Lee:2010ik,Gehrmann:2010ue,Gehrmann:2010tu,vonManteuffel:2015gxa})
and only very recently the complete four loop results became
available~\cite{Lee:2021uqq,LMSSSS22}. Massive quark form factors are known at
two-loop order from
Refs.~\cite{Mastrolia:2003yz,Bonciani:2003ai,Bernreuther:2004ih,Gluza:2009yy,Henn:2016tyf,Ahmed:2017gyt,Ablinger:2017hst,Lee:2018nxa}. At
three loops only partial results are available, namely all planar
contributions needed for the large-$N_c$ limit (where $N_c$ is the number of
colours in QCD)~\cite{Henn:2016tyf,Ablinger:2018yae} and the fermionic
contributions with closed massless quark loops~\cite{Lee:2018nxa}. For the
contribution involving massive closed fermion loops a deep expansion with at
least 2000 terms around the on-shell photon limit has been computed in
Ref.~\cite{Blumlein:2019oas}.

The available results show an involved analytic structure
containing iterated integrals with the letters
$x$, $1-x$, $1+x$ and $x-e^{i\pi/3}$, where the relation between $x$
and the photon virtuality $s=q^2$ is given by
\begin{eqnarray}
  \frac{q^2}{m^2} &=& - \frac{(1-x)^2}{x}\,,
                      \label{eq::xx}
\end{eqnarray}
with $m$ the mass of the heavy quark.  A numerical evaluation of the analytic
expressions is possible using, e.g., {\tt
  ginac}~\cite{Bauer:2000cp,Vollinga:2004sn}.  However, depending on the phase
space point it might be time consuming and/or its numerical accuracy is
limited to a few digits only.  Thus, in practice, one often resolves to the
construction of approximations which enable a fast numerical evaluation.
Moreover, the three-loop results for the colour structures which are not yet
available in analytic form cannot be expressed in terms of simple iterated
integrals. Rather, so-called elliptic integrals are present as the
fundamental building blocks.  Currently there is no ready-to-use approach for
the numerical evaluation of the corresponding mathematical functions and thus
especially here numerical approximations are needed.

\begin{figure}[t]
  \begin{center}
    \begin{tabular}{ccc}
      \includegraphics[width=0.15\textwidth]{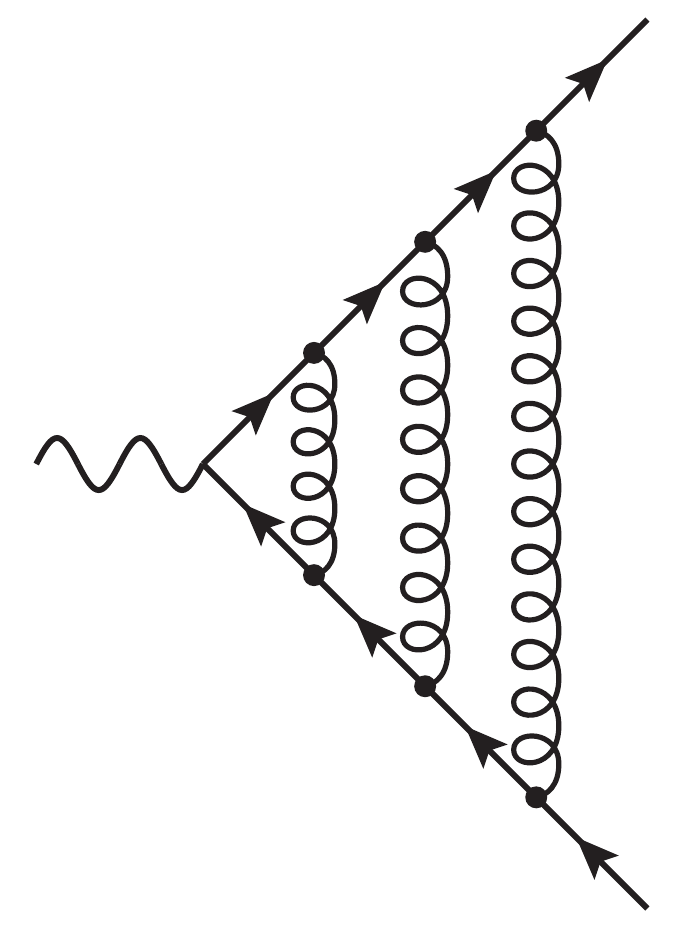} &
      \includegraphics[width=0.15\textwidth]{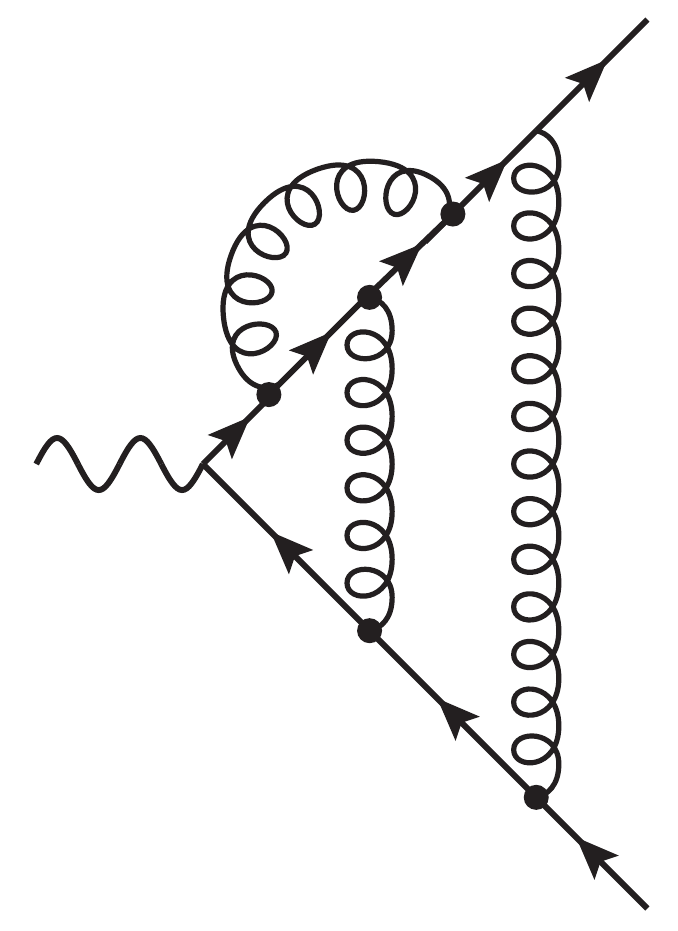} &
      \includegraphics[width=0.15\textwidth]{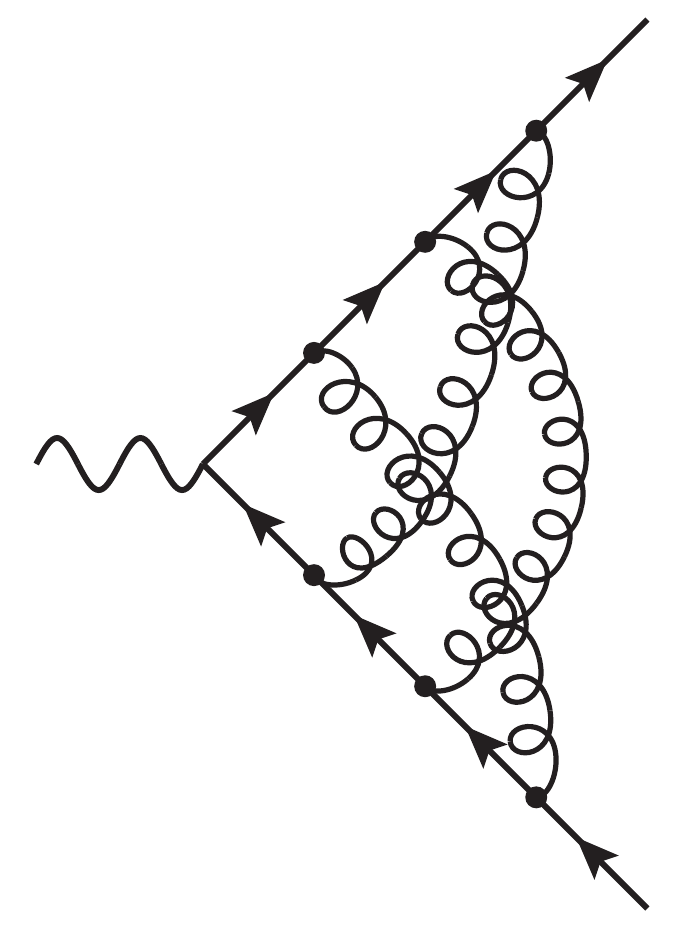}
      \\
      (a) & (b) & (c)
    \end{tabular}
    \caption{\label{fig::diags}Sample Feynman diagrams for the vector form
      factors at three loops. Solid and curly lines denote quarks and gluons,
      respectively. The external photon is represented by a wavy line.}
  \end{center}
\end{figure}

In this Letter we present results for the three-loop  form
factor with an external vector current. We consider QCD with one massive and
$n_l$ massless flavours and compute the
non-singlet contribution, where the external quarks directly couple to the
current, see also the sample Feynman diagrams in Fig.~\ref{fig::diags}.
We perform the reduction to master
integrals and establish the differential equations for the latter. They are
used in order to construct expansions around singular and regular points using
analytic results at $s=0$ as initial condition.
In our calculation we keep the symbols for the Casimir operators of SU$(N_c)$
and thus obtain results for each individual colour factor.

There are other methods which are based on difference or differential
equations accompanied by
expansions~\cite{Laporta:2001dd,Boughezal:2007ny,Blumlein:2017dxp,Liu:2017jxz,Lee:2017qql,Moriello:2019yhu,Hidding:2020ytt,Dubovyk:2022frj}. However,
some of them have only been applied to individual master integrals and they
are still lacking the proof that they can handle non-trivial
physical problems with a few hundred master integrals.
In this paper we apply the method of Ref.~\cite{Fael:2021kyg} to a non-trivial
physical quantity and show that numerically precise results can be obtained in
the whole parameter space.


\bigskip {\bf Calculation.}
We consider the photon-quark vertex and define the Dirac and Pauli
form factors as
\begin{eqnarray}
  \Gamma_\mu(q_1,q_2) &=&
  F_1(q^2)\gamma_\mu - \frac{i}{2m}F_2(q^2) \sigma_{\mu\nu}q^\nu
  \,,
  \label{eq::F1F2}
\end{eqnarray}
with incoming momentum $q_1$, outgoing momentum $q_2$ and $q=q_1-q_2$.  The
external quarks are on-shell and we have
$\sigma^{\mu\nu} = i[\gamma^\mu,\gamma^\nu]/2$. The colour factor is a simple
Kronecker delta in the fundamental colour indices of the external quarks and
it is suppressed for convenience.  $F_1$ and $F_2$ can easily be obtained by applying
appropriate projectors.

Sample Feynman diagrams are shown in Fig.~\ref{fig::diags}.  We generate the
amplitudes with {\tt qgraf}~\cite{Nogueira:1991ex} and use {\tt q2e} and~{\tt
  exp}~\cite{Harlander:1997zb,Seidensticker:1999bb,q2eexp} to rewrite the
output to {\tt FORM}~\cite{Kuipers:2012rf} notation and map each diagram to a
predefined integral family.  In this way we can express $F_1$ and $F_2$ as a
linear combination of scalar functions with twelve indices where nine
correspond to the exponents of propagators and the remaining
three to the exponents of
irreducible numerators.

For each integral family we use {\tt
  Kira}~\cite{Maierhofer:2017gsa,Klappert:2020nbg} with {\tt Fermat}~\cite{fermat}
to reduce the scalar
functions to master integrals. In this step we take care to choose a good
basis such that for each entry in our integral tables the dependence on the
space-time dimension $d=4-2\epsilon$ and the kinematic variables $s$ and $m^2$
factorizes in the denominators. This is done with the help of an improved
version of the program developed in Ref.~\cite{Smirnov:2020quc}.  {\tt Kira}
is also used to minimize the number of master integrals over all
families. This allows us to express $F_1$ and $F_2$ in terms of 422 master
integrals.

In a next step we establish differential equations for the
master integrals using {\tt LiteRed}~\cite{Lee:2012cn,Lee:2013mka} and {\tt
  Kira}
and use the results for $s\to0$ as initial conditions.
In fact, the construction of the solution can be organized such that
the naive limit $s=0$ of a subset of the 422 master integrals
is sufficient to fix all unknown constants.

In the limit $s=0$ the vertex integrals reduce to two-point on-shell
integrals, which have been studied in
Refs.~\cite{Laporta:1996mq,Melnikov:2000qh}.  We use the results for the
corresponding master integrals from Ref.~\cite{Lee:2010ik} which are available
up to weight~7. Due to spurious poles in $\epsilon$ some of the
on-shell master integrals are needed to higher weight which can be constructed
with the help of Ref.~\cite{Lee:2015eva} and PSLQ~\cite{PSLQ}
(see also Ref.~\cite{Blumlein:2019oas}). For the current calculation a subset
of integrals was needed up to weight~9.

After fixing the initial conditions we can use the differential equations to
obtain for each master integral an expansion in $s/m^2$ up to
$(s/m^2)^{75}$. For all other expansions described below we have computed 50
expansion terms. In this context the use of finite fields with a special
version of {\tt Kira} and
{\tt FireFly}~\cite{Klappert:2019emp,Klappert:2020aqs} was essential for our
calculation.  Starting from $s=0$ we move both to negative and positive values
of $s$.  To do so we choose values $s_0/m^2=1$ and $s_0/m^2=-4$ and
construct generic expansions with the help of the differential equations. They
are matched to the $s=0$ expansion by evaluating the latter numerically at
$s/m^2=1/2$ and $s/m^2=-2$, respectively.  This provides initial conditions
for the $s_0$ expansions.  In total we construct expansions around the
following 30 values\footnote{Note that only one expansion for large
  absolute values of $s$ is necessary to cover the limits $s\to\pm \infty$.}
\begin{eqnarray}
  \frac{s_0}{m^2}\!\in\!\{-\infty,-32,-28,-24,-16,-12,-8,-4,0,1,2,5/2,
  \nonumber\\
  \!\!3,7/2,4,9/2,5,6,7,8,10,12,14,15,16,17,19,22,28,40\}
  \nonumber\\
                                       \label{eq::s0}
\end{eqnarray}
and perform the matching step-by-step starting from $s=0$. In this way we
can cover the whole $s/m^2$ plane. For more details on the ``expansion and
matching'' method we refer to Ref.~\cite{Fael:2021kyg}.

At first sight it seems that the variable $x$ introduced in
Eq.~(\ref{eq::xx}) is the proper variable to perform the expansions, since
the characteristic points $s/m^2 = 0,4,\infty$ correspond to $x=1,-1,0$.
However, in practice it is more advantageous to work in $s/m^2$.
This is also connected to the new threshold at $s/m^2 = 16$
which appears for the first time at three loops. It is mapped to
$x=4\sqrt{3}-7\approx-0.072$ which limits the radius of convergence of the
variable $x$.

Let us in the following comment on the choice of $s_0$ in Eq.~(\ref{eq::s0}).
Some values correspond to a particular kinematic situation:
$s/m^2=4$ and $16$ correspond to the two- and four-particle thresholds and
$m^2/s=0$ to the high energy limit. Furthermore, as mentioned above, we
compute the initial conditions for $s=0$.
To guarantee sufficient accuracy over the whole $s/m^2$ range we have
introduced further expansions for positive and negative values of $s$.
In the differential equations we
observe further singularities for $s/m^2\in\{-4,-2,-1,-1/2,1/2,1,2,3,16/3\}$.
However, they are spurious since the form factors are regular for these
values of $s$.
Nevertheless, for some of them we have constructed an
expansion of the master integrals.

For all expansions the convergence around a given value $s_0$ is only
guaranteed up to the next singular point in the complex $s$ plane.  For
example for $s_0/m^2=22$ we have convergence for $16<s/m^2<28$ and for
$s_0/m^2=-4$ for $-12<s/m^2<4$.  Note that $s/m^2=4,16$ and $\infty$ are
singular points of the differential equation which require a power-log
expansion. Furthermore, for $s/m^2=4$ and $16$ we have an expansion
in $\sqrt{4-s/m^2}$ and $\sqrt{16-s/m^2}$, respectively.  For all other points
simple Taylor expansions are sufficient.

Often the convergence of a series expansion can be enhanced
by switching to a different expansion parameter.
One powerful method is based on M\"obius transformations
as has already been discussed in Ref.~\cite{Lee:2017qql}.
Assume, we want to expand around the point $x_k$ and there
are singular points of the differential equations at $x_{k-1}$
and $x_{k+1}$ with $x_{k-1} < x_{k} < x_{k+1}$.
Naively the radius of convergence is limited by the distance
to the closer singular point.
However, the variable transformation
\begin{align}
  y_k =
  \frac{(x-x_k)(x_{k+1}-x_{k-1})}{(x-x_{k+1})(x_{k-1}-x_{k})+(x-x_{k-1})(x_{k+1}-x_k)}
\end{align}
maps the points $x_{k-1}$, $x_{k}$, $x_{k+1}$ to $-1$, $0$, $1$.  The reach of
the series expansion is therefore extended in the direction of the
farthest singularity although the convergence at the boundaries can be
quite slow.  We find this mapping indispensable when constructing regular
series expansions close to singular points.

The form factors $F_1$ and $F_2$ develop both ultraviolet and infrared
divergences. The former are taken care of by counterterms for the wave
functions and mass of the heavy quarks, which we renormalize on-shell.
Furthermore, the strong coupling constant is renormalized in the
$\overline{\rm MS}$ scheme. The remaining infrared poles are described by a
universal function independent of the external current, the cusp anomalous
dimension $\Gamma_{\rm cusp}$, which has been computed to three-loop accuracy
in Refs.~\cite{Grozin:2014hna,Grozin:2015kna}.
It is used to
construct a $Z$ factor (see, e.g., Ref.~\cite{Lee:2018rgs}) such that
the combination
\begin{eqnarray}
  F_{1,2} = Z F_{1,2}^{f}
  \label{eq::F_Ff}
\end{eqnarray}
leads to the ultraviolet and infrared finite form factors
$F_{1,2}^{f}$. We introduce their perturbative expansion as
\begin{eqnarray}
  F_{1,2}^{f} &=& \sum_{n\ge0} F_{1,2}^{f,(n)} \left(\frac{\alpha_s}{\pi}\right)^n
                  \,,
                  \label{eq::F12exp}
\end{eqnarray}
where $F_{1}^{f,(0)}=1$ and $F_{2}^{f,(0)}=0$.  Since $Z$ is expressed in
terms of the strong coupling in the effective $n_l$-flavour theory we have
$\alpha_s\equiv\alpha_s^{(n_l)}(\mu)$ in Eq.~(\ref{eq::F12exp}).  In the next
Section we discuss results for $F_1^{f,(3)}$ and $F_2^{f,(3)}$.


\bigskip {\bf Results.}  The results from our calculation are
expansions around the values $s_0$ in Eq.~(\ref{eq::s0}).  Thus, we can define
the form factors $F_1$ and $F_2$ piecewise by these expansions.
We choose for the renormalization scale $\mu^2=m^2$.

\begin{figure}[b]
  \begin{center}
    \begin{tabular}{cc}
      \includegraphics[width=0.23\textwidth]{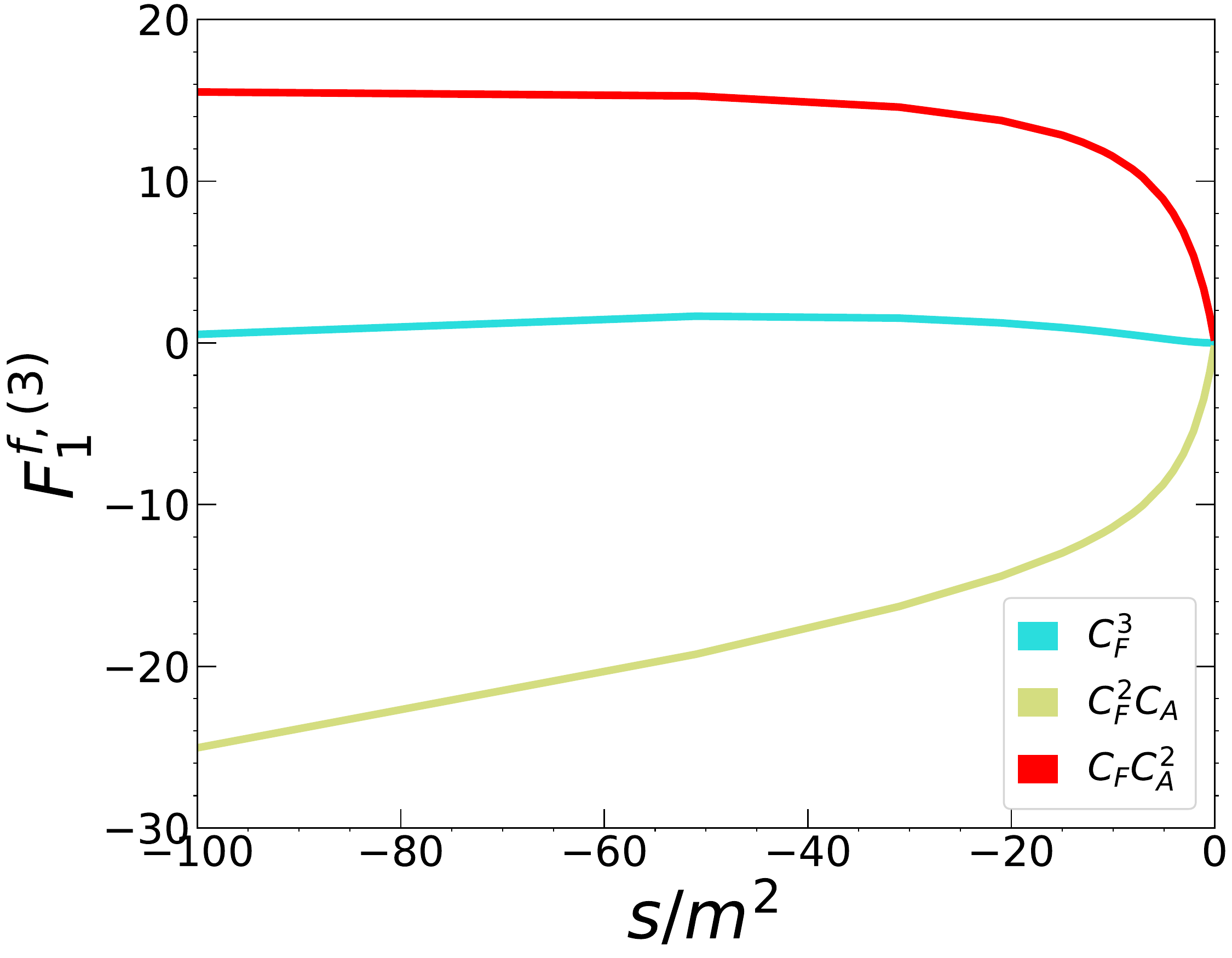}
      &
      \includegraphics[width=0.23\textwidth]{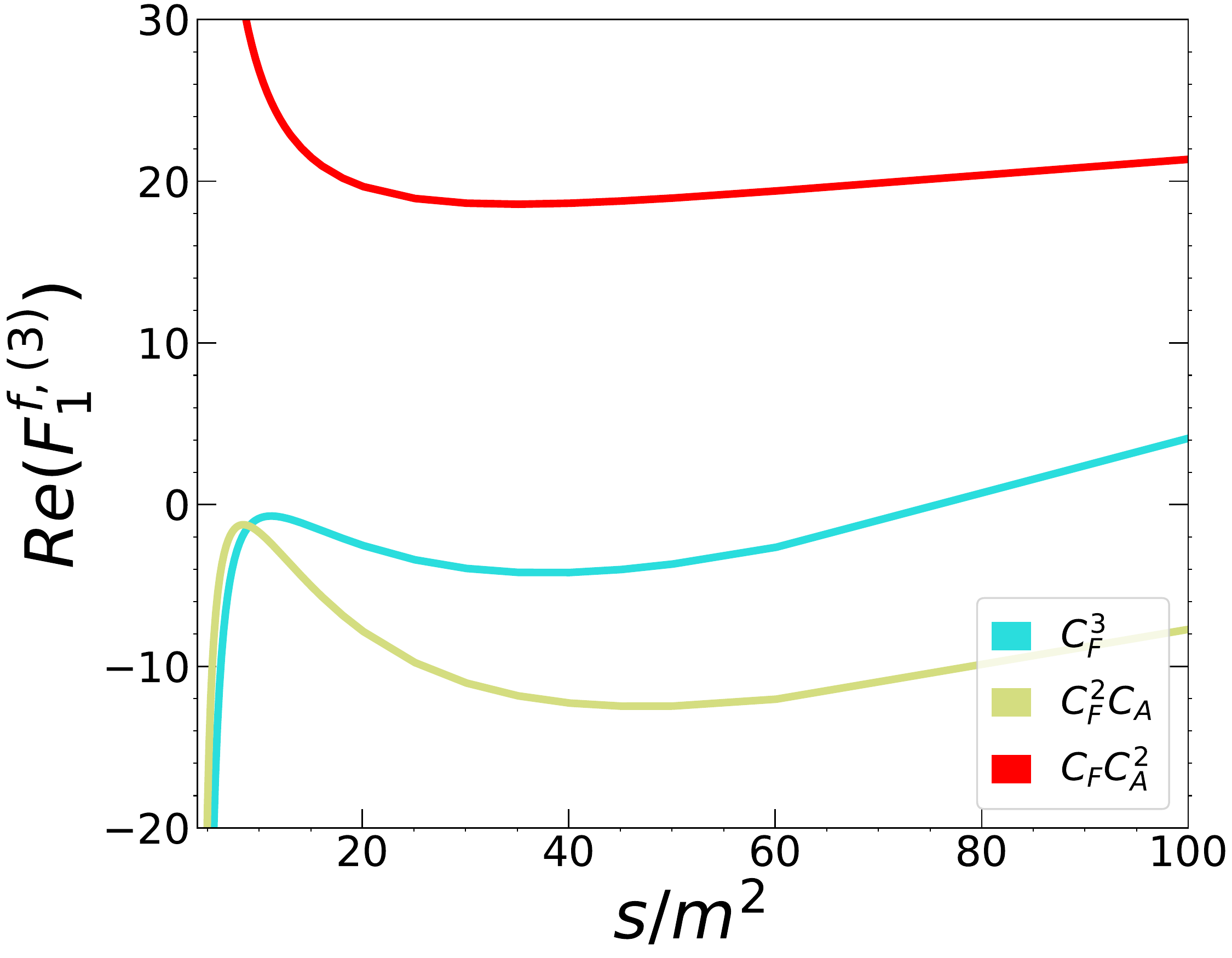}
      \\
      (a) & (b)
    \end{tabular}
    \caption{\label{fig::F1}The colour structures $C_F^3$, $C_F^2 C_A$, $C_F
      C_A^2$ of $F_1^f$ as a function of $s$. We show results
      for $s<0$ (a) and $s>4m^2$ (b).}
  \end{center}
\end{figure}

In the following we concentrate on $F_1$ and present results for
the renormalized and infrared-subtracted form factor.
In Fig.~\ref{fig::F1} we illustrate the results for the
three non-fermionic colour structures $C_F^3$, $C_F^2 C_A$, $C_FC_A^2$,
where $C_F$ and $C_A$ are the Casimir operators of the
fundamental and the adjoint representation, respectively,
and present results for $s<0$ and $s>4m^2$. For $s=0$ we have $F_1=0$
as can be seen in plot~(a). In plot~(b) one observes the
influence of the Coulomb singularity even for $s/m^2\approx 10$.
The four-particle threshold is much less pronounced.
In the high-energy region, both for $s>0$ and $s<0$
the form factor contains logarithms up to sixth order.

We estimate the accuracy of our result from the numerical pole cancellations
of the renormalized and infrared subtracted form factor.  For $s>4m^2$ the
quadratic and linear $1/\epsilon$ poles cancel with a relative precision of
$10^{-12}$ and $10^{-10}$, respectively.  Assuming a similar progression we
estimate that for the finite term we have at least eight significant digits
for the coefficients of each colour factor.  In the regions $0<s<4m^2$ and
$s<0$ the accuracy is significantly higher and in general exeeds twelve
significant digits.  Also for the fermionic colour structures a notably higher
accuracy is reached.

In a next step we consider the special kinematic points $s=0, 4m^2, 16 m^2$
and $\pm\infty$ and present (numerical) expansions using the genuine results
of our approximation methods.  In this Letter we restrict
ourselves to the non-fermionic colour factors.  In the supplemenatry material we
present results for the contributions which contain a closed heavy quark
loop. The remaining fermionic contributions are available in the
literature~\cite{Lee:2018nxa,Lee:2018rgs}.

In the static limit we construct an analytic expansion up to
$s^{67}$ from the boundary values at $s=0$. The first two expansion
terms are given by
\newcommand{\logtwo}{l_2}
\begin{widetext}
\begin{eqnarray}
  \lefteqn{F_1^{f,(3)}\Big|_{s\to0} = \Bigg\{
  C_A C_F^2
  \Bigg[
    \frac{19 a_4}{2}
    -\frac{\pi ^2 \zeta _3}{9}
    +\frac{17725 \zeta _3}{3456}
    -\frac{55 \zeta _5}{32}
    +\frac{19 l_2^4}{48}
    -\frac{97}{720} \pi ^2 l_2^2
    +\frac{29 \pi ^2 l_2}{240}
    -\frac{347 \pi ^4}{17280}
    -\frac{4829 \pi ^2}{10368}
    +\frac{707}{288}
  \Bigg]}
\nonumber\\&&\mbox{}
  +C_A^2 C_F
  \Bigg[
    -a_4
    +\frac{7 \pi ^2 \zeta _3}{96}
    +\frac{4045 \zeta _3}{5184}
    -\frac{5 \zeta _5}{64}
    -\frac{l_2^4}{24}
    +\frac{67}{360} \pi ^2 l_2^2
    -\frac{5131 \pi ^2 l_2}{2880}
    +\frac{67 \pi ^4}{8640}
    +\frac{172285 \pi ^2}{186624}
    -\frac{7876}{2187}
  \Bigg]
\nonumber\\&&\mbox{}
  +C_F^3
  \Bigg[
    -15 a_4
    -\frac{17 \pi ^2 \zeta _3}{24}
    -\frac{18367 \zeta _3}{1728}
    +\frac{25 \zeta _5}{8}
    -\frac{5 l_2^4}{8}
    -\frac{19}{40} \pi ^2 l_2^2
    +\frac{4957 \pi ^2 l_2}{720}
    +\frac{3037 \pi ^4}{25920}
    -\frac{24463 \pi ^2}{7776}
    +\frac{13135}{20736}
  \Bigg]\Bigg\}\frac{s}{m^2}
                                       \nonumber\\&&\mbox{}
                                        + {\cal O}\left(\frac{s^2}{m^4}\right)
                                        +\mbox{fermionic contributions}\,,
                               \label{eq::F1s0}
\end{eqnarray}

\vspace*{-1.5em}

\end{widetext}
where $l_2=\log(2)$, $a_4=\mbox{Li}_4(1/2)$ and $\zeta_n$ is Riemann's zeta function
evaluated at $n$.

The first two terms for the high-energy expansion
of the non-fermionic colour structures read
\newcommand{\cRthree}{C_F^3}
\newcommand{\cRtwocA}{C_F^2C_A}
\newcommand{\cRcAtwo}{C_FC_A^2}
\newcommand{\logzero}{}
\newcommand{\logone}{l_s}
\renewcommand{\logtwo}{l_s^2}
\newcommand{\logthree}{l_s^3}
\newcommand{\logfour}{l_s^4}
\newcommand{\logfive}{l_s^5}
\newcommand{\logsix}{l_s^6}
\begin{widetext}
\begin{eqnarray}
 \lefteqn{  F_1^{f,(3)}\Big|_{s\to-\infty} =
   4.7318\cRthree-20.762\cRtwocA+ 8.3501\cRcAtwo
+ \Big[ 3.4586\cRthree-4.0082\cRtwocA-6.3561\cRcAtwo \Big]\logone}
\nonumber\\ && \mbox{}
+ \Big[ 1.4025\cRthree+ 0.51078\cRtwocA -2.2488\cRcAtwo \Big]\logtwo
+ \Big[ 0.062184\cRthree+ 0.90267\cRtwocA-0.42778\cRcAtwo \Big]\logthree
\nonumber\\ && \mbox{}
+ \Big[-0.075860\cRthree+0.20814\cRtwocA-0.035011\cRcAtwo \Big]\logfour
+ \Big[{-0.023438}\cRthree+0.019097\cRtwocA \Big]\logfive
\nonumber\\ && \mbox{}
+ \Big[{-0.0026042}\cRthree \Big]\logsix
+ \Big\{
-92.918\cRthree+123.65\cRtwocA-47.821\cRcAtwo
+ \Big[-10.381\cRthree+2.3223\cRtwocA
\nonumber\\ && \mbox{}
               +17.305\cRcAtwo \Big]\logone
+ \Big[ 4.9856\cRthree-19.097\cRtwocA+8.0183\cRcAtwo \Big]\logtwo
+ \Big[ {3.0499}\cRthree {-6.8519}\cRtwocA+  1.9149\cRcAtwo \Big]\logthree
\nonumber\\ && \mbox{}
+ \Big[ 0.67172\cRthree -0.91213\cRtwocA+  0.24069\cRcAtwo \Big]\logfour
+ \Big[ 0.13229\cRthree-0.051389\cRtwocA+ 0.0043403\cRcAtwo \Big]\logfive
\nonumber\\ && \mbox{}
+ \Big[ 0.0041667\cRthree-0.0010417\cRtwocA- 0.00052083\cRcAtwo \Big]\logsix
\Big\} \frac{m^2}{s}
+ {\cal O}\left(\frac{m^4}{s^2}\right)
+ \mbox{fermionic contributions}\,,
                               \label{eq::F1sminf}
\end{eqnarray}
\end{widetext}
with $l_s = \log(m^2/(-s-i\delta))$. The leading logarithmic contributions of the order
$\alpha_s^n \log^{2n}(m^2/s)$ are given by the Sudakov
exponent~\cite{Sudakov:1954sw,Frenkel:1976bj}
$\mbox{exp}[-C_F\alpha_s/(4\pi) \times \log^2 (m^2/s)]$ which is reproduced by
our expansions.  In fact, in our calculation we can even reconstruct the
analytic results of the coefficients which are given by
\begin{eqnarray}
F_1^{f,{(3)}} \!\!=\! -\frac{C_F^3}{384} l_s^6
                                 + \frac{m^2}{s}\left( \frac{C_F^3}{240}
     - \frac{C_F^2C_A}{960} - \frac{C_FC_A^2}{1920}\right) l_s^6
  \!+ \ldots
                  \nonumber\\
                  \label{eq::F1sminf_lead}
\end{eqnarray}
In Eq.~(\ref{eq::F1sminf}) they are shown in numeric form.  Note that also the
leading logarithms of the mass corrections $m^2/s$ perfectly agree with
Ref.~\cite{Liu:2017axv} where the results in Eq.~(\ref{eq::F1sminf_lead}) have
been obtained using an involved asymptotic expansion of the three-loop vertex
diagrams.  Our approach provides the whole tower of logarithms and also higher
order $m^2/s$ contributions. We estimate the accuracy of the non-logarithmic
term in Eq.~(\ref{eq::F1sminf}) to ten digits. For the subleading terms
the accuracy decreases. Note, however, that we use the $s\to\infty$ expansion
only for $|s/m^2| \gtrsim 45$ and that
$1/45^3 \approx {\cal O}(10^{-5})$.

Let us next discuss the thresholds at $s=4m^2$ and $s=16m^2$.
Close to the two-particle threshold $F_1$ develops the famous Coulomb
singularity with negative powers in the velocity of the produced quarks,
$\beta = \sqrt{1 - 4 m^2/s}$, up to
third order multiplied by $\log(\beta)$ terms.
Close to threshold it is interesting to consider the combination of $F_1$ and $F_2$
\begin{eqnarray}
  \frac{3}{2}\Delta&=& |F_1 + F_2|^2
    + \frac{ |(1-\beta^2) F_1 + F_2 |^2}{2(1-\beta^2)}
  \,,
\end{eqnarray}
which is closely related to the cross section of heavy quark production in
electron positron annihilation via
$\sigma(e^+e^-\to Q \bar{Q})=\sigma_0  \beta  3\Delta/2$ with
$\sigma_0 = 4\pi \alpha^2 Q_Q^2 /(3s)$, where $\alpha$ is the fine structure
constant and $Q_Q$ is the fractional charge of the massive quark $Q$. For
$\beta\to0$ real radiation is suppressed by two powers of $\beta$ which allows
us to provide the first two terms in the expansion for each colour factor. Our
result for the third order correction $\Delta^{(3)}$ reads
\begin{widetext}
\begin{align}
  & \Delta^{(3)} =
   C_F^3 \bigl[
    -\frac{32.470}{\beta ^2}
    + \frac{1}{\beta } \bigl( 14.998-32.470 l_{2\beta} \bigr)
  \bigr]
  + C_A^2 C_F \frac{1}{\beta} \bigl[ 16.586 l^2_{2 \beta}-22.572 l_{2\beta}+42.936\bigr]
\nonumber\\&
  + C_A C_F^2 \bigl[
    \frac{1}{\beta^2} \bigl( -29.764 l_{2\beta}-7.770339\bigr)
    +\frac{1}{\beta} \bigl( -12.516 l_{2 \beta}-11.435 \bigr)
  \bigr]
  + \mathcal{O}(\beta^0)
               + \mbox{fermionic contributions}\,,
\end{align}
\end{widetext}
with $l_{2\beta}=\log(2\beta)$.
Our numerical results reproduce the analytic expressions
from Ref.~\cite{Kiyo:2009gb} (see also
Refs.~\cite{Pineda:2006ri,Hoang:2008qy})
with at least 13 digits accuracy.

Four-particle thresholds are present in diagrams which contain a closed heavy
quark loop but also in purely gluonic diagrams like the one in
Fig.~\ref{fig::diags}(b).  Interestingly  it has a smooth
behaviour. In fact, we observe the first non-analytic terms at order
$(\beta_4)^9$ with $\beta_4 = \sqrt{1 - 16 m^2/s}$.  Note that the massive
four-particle phase-space, which is one of our master integrals, already
provides a factor $(\beta_4)^7$. Furthermore, our
expansions of $F_1$ and $F_2$ up to $(\beta_4)^{50}$ do not contain any
$\log\beta_4$ terms although many of the master integrals contain such terms.

Finally, we want to mention that we
have performed the calculation for general QCD gauge parameter
$\xi$ and have checked that $\xi$ cancels in the renormalized form factors.
Note that both the bare three-loop expressions and the quark mass
counterterm contributions depend on $\xi$.
Furthermore, we can specify our result to the large-$N_c$
limit and compare against the exact results from Ref.~\cite{Henn:2016tyf}. In
this limit only about 90 planar master integrals contribute and we observe a
significantly increased precision of our result. In fact, in the whole $s/m^2$
region we can reproduce the exact result with at least 14 digits.


\bigskip {\bf Conclusions.}  In this Letter we present for the first time
results for the non-singlet three-loop massive photon-quark form factors taking
into account all colour structures. We use the methods based on ``expansion
and matching'' as introduced in Ref.~\cite{Fael:2021kyg} and obtain numerical
approximations in the whole $s/m^2$ range.  Based on the comparison to
the partially known exact results and on internal cross checks of the method
we estimate the accuracy to at least eight significant digits above the
$s=4m^2$ threshold and to about twelve digits below. Note that, if required, a
systematic improvement is possible by adding more intermediate matching
points. The application to a physical quantity with a non-trivial analytic
structure shows the effectiveness of our method.


\smallskip

{\bf Acknowledgements.}  We thank Roman Lee for discussions about the M\"obius
transformations and Alexander Smirnov and Vladimir Smirnov for discussions
about the basis change for the master integrals and providing an improved
version of the {\tt Mathematica} code from Ref.~\cite{Smirnov:2020quc}.  This
research was supported by the Deutsche Forschungsgemeinschaft (DFG, German
Research Foundation) under grant 396021762 --- TRR 257 ``Particle Physics
Phenomenology after the Higgs Discovery''.  The Feynman diagrams were drawn
with the help of Axodraw~\cite{Vermaseren:1994je} and
JaxoDraw~\cite{Binosi:2003yf}.

\newpage

\begin{appendix}

\begin{widetext}
\section*{Supplementary material}

To complete the presentation of the main part
of the Letter we provide in the following results for
$F_1$ involving a closed heavy quark for $s\to0$, $s\to-\infty$
and $s\to 4m^2$.
For such contributions one often introduces the tag
$n_h$, which means that we present result for the
colour factors $C_F^2 T_F n_h$, $C_F C_A T_F n_h$,
$C_F T_F^2 n_h^2$ and $C_F T_F^2 n_h n_l$.

For $s\to0$ we have
\begin{eqnarray}
  \lefteqn{F_1^{f,(3),n_h}\Big|_{s\to0} = \Bigg\{
  C_A C_F T_F n_h
  \Bigg(\frac{17 a_4}{6}
  -\frac{\pi ^2 \zeta _3}{18}
  +\frac{1775 \zeta _3}{864}
  +\frac{5 \zeta _5}{12}
  +\frac{17 l_2^4}{144}
  -\frac{17}{144} \pi ^2 l_2^2
  -\frac{149 \pi ^2 l_2}{108}
  +\frac{803 \pi ^4}{51840}
}
\nonumber\\&&\mbox{}
  +\frac{4813 \pi ^2}{5184}
  -\frac{23089}{5184}
\Bigg)
+C_F^2 T_F n_h
\Bigg(
  -\frac{32 a_4}{9}
  +\frac{1441 \zeta _3}{1728}
  -\frac{4 l_2^4}{27}
  +\frac{4}{27} \pi ^2 l_2^2
  -\frac{2 \pi ^2 l_2}{9}
  -\frac{13 \pi ^4}{1620}
  +\frac{1057 \pi ^2}{2430}
  -\frac{2273}{1296}
\Bigg)
\nonumber\\&&\mbox{}
+C_F T_F^2 n_h n_l
\Bigg(
  \frac{7 \zeta _3}{48}
  -\frac{1}{24} \pi ^2 l_2
  -\frac{71 \pi ^2}{1296}
  +\frac{1261}{1944}
\Bigg)
+ C_F T_F^2 n_h^2
\Bigg(
  -\frac{28 \zeta _3}{27}
  +\frac{4 \pi ^2}{135}
  +\frac{11257}{11664}
\Bigg)\Bigg\}\frac{s}{m^2} + {\cal O}\left(\frac{s^2}{m^4}\right)\,.
\end{eqnarray}

In the high-energy limit the $n_h$ contributions are given by
\newcommand{\cRnhtwo}{C_FT_F^2n_h^2}
\newcommand{\cRnhnl} {C_FT_F^2n_hn_l}
\newcommand{\cRtwonh}{C_F^2T_Fn_h}
\newcommand{\cRcAnh} {C_FC_AT_Fn_h}
\begin{eqnarray}
  \lefteqn{F_1^{f,(3),n_h}\Big|_{s\to-\infty} =
    -1.54208\cRnhtwo -4.1144\cRnhnl -3.2872\cRtwonh+  10.425\cRcAnh}
\nonumber\\ && \mbox{}
+ \Big[   -1.2844\cRnhtwo -2.8537\cRnhnl -2.8785\cRtwonh+   7.6917\cRcAnh \Big]\logone
\nonumber\\ && \mbox{}
+ \Big[   -0.40466\cRnhtwo-0.80931\cRnhnl-1.8900\cRtwonh + 2.2962\cRcAnh \Big]\logtwo
\nonumber\\ && \mbox{}
+ \Big[   -0.058642\cRnhtwo-0.11728\cRnhnl-0.55727\cRtwonh + 0.33008\cRcAnh \Big]\logthree
\nonumber\\ && \mbox{}
+ \Big[   -0.0046296\cRnhtwo-0.0092593\cRnhnl-0.086806\cRtwonh + 0.025463\cRcAnh \Big]\logfour
\nonumber\\ && \mbox{}
+ \Big[   -0.0069444\cRtwonh \Big]\logfive
+ \Big\{
    11.898\cRnhtwo+  18.981\cRnhnl-5.2612\cRtwonh
+ -52.115\cRcAnh 
\nonumber\\ && \mbox{}
+ \Big[    7.2323\cRnhtwo + 8.7158\cRnhnl + 3.3633\cRtwonh+ -25.912\cRcAnh \Big]\logone
\nonumber\\ && \mbox{}
+ \Big[    1.8056\cRnhtwo + 2.5000\cRnhnl + 8.4570\cRtwonh-7.8739\cRcAnh \Big]\logtwo
+ \Big[    0.27778\cRnhtwo
\nonumber\\ && \mbox{}
 + 0.33333\cRnhnl + 2.3758\cRtwonh-1.4464\cRcAnh \Big]\logthree
+ \Big[    0.48843\cRtwonh
\nonumber\\ && \mbox{}
-0.067130\cRcAnh \Big]\logfour
+ \Big[    0.0069444\cRtwonh-0.0034722\cRcAnh \Big]\logfive
\Big\} \frac{m^2}{s} +  {\cal O}\left(\frac{m^4}{s^2}\right) \,.
\end{eqnarray}

For $\beta\to0$ the results read
\begin{eqnarray}
  \Delta^{(3),n_h} &=&
    C_F^2 T_F n_h \bigl[ \frac{2.4792}{\beta }-1.3159 l_{2\beta}-2.0339 \bigr]
  - C_A C_F T_F n_h \bigl[ 0.082247 \beta  + 0.20495 \bigr]
  + 0.10248 C_F T_F^2 n_h^2
  \nonumber\\&&\mbox{}
  + C_F T_F^2 n_h n_l \bigl[ 0.87730 \beta -0.54050 \bigr]
  + \ldots\,,
\end{eqnarray}
where the ellipses denote higher order terms in $\beta$.

\end{widetext}

\end{appendix}

\end{document}